\documentclass[prl,twocolumn,showpacs,floats]{revtex4}

\usepackage{graphicx}

\begin{document}

%
\title{High Energy Dispersions in Bi$_2$Sr$_2$CaCu$_2$O$_8$ High Temperature Superconductor from Laser-Based Angle-Resolved Photoemission}
%
%
%
\author{Wentao Zhang$^{1}$, Guodong Liu$^{1}$, Jianqiao Meng$^{1}$, Lin Zhao$^{1}$, Haiyun Liu$^{1}$, Xiaoli Dong$^{1}$, Wei Lu$^{1}$, J. S. Wen$^{2}$, Z. J. Xu$^{2}$, G. D. Gu$^{2}$, T. Sasagawa$^{3}$, Guiling Wang$^{4}$, Yong Zhu$^{5}$, Hongbo Zhang$^{4}$,Yong Zhou$^{4}$, Xiaoyang Wang$^{5}$, Zhongxian Zhao$^{1}$, Chuangtian Chen$^{5}$, Zuyan Xu$^{4}$ and X. J. Zhou $^{1,*}$}

\affiliation{
\\$^{1}$National Laboratory for Superconductivity, Beijing National Laboratory for Condensed Matter Physics, Institute of Physics, Chinese Academy of Sciences, Beijing 100080, China
\\$^{2}$Condensed Matter Physics and Materials Science Department, Brookhaven National Laboratory, Upton, New York 11973, USA
\\$^{3}$Materials and Structures Laboratory, Tokyo Institute of Technology, Yokohama Kanagawa, Japan
\\$^{4}$Laboratory for Optics, Beijing National Laboratory for Condensed Matter Physics,Institute of Physics, Chinese Academy of Sciences, Beijing 100080,  China
\\$^{5}$Technical Institute of Physics and Chemistry, Chinese Academy of Sciences, Beijing 100080, China}
\date{January 16, 2008}
%
%

\begin{abstract}

Super-high resolution laser-based angle-resolved photoemission (ARPES) measurements have been carried out on the high energy electron dynamics in Bi$_2$Sr$_2$CaCu$_2$O$_8$ (Bi2212) high temperature superconductor. Momentum dependent measurements provide new insights on the nature of high energy kink at 200$\sim$400 meV and high energy dispersions. The strong dichotomy between the MDC- and EDC-derived bands is revealed which raises critical issues about its origin and which one represents intrinsic band structure.  The MDC-derived high energy features are affected by the high-intensity valence band at higher binding energy and may not be intrinsic.

\end{abstract}


\maketitle

%

High temperature superconductivity in cuprates is realized by doping charge carriers into the CuO$_2$ planes where the oxygen 2{\it p} orbitals and copper 3{\it d} orbitals hybridize to dictate the basic electronic structure of these materials. It is also well-known that the unusual physical properties of high-T$_C$ cuprates stem from the complex interplay between electron charge, spin and lattice vibrations.  Reveling basic electronic structure and probing many-body effects are essential for understanding the anomalous phenomena in high temperature superconudtcors\cite{ThreeReviews}.
Very recently, great effort has been focused on identifying high energy scales in high-T$_c$ superconductors\cite{Ronning,Graf,Xie,Valla,Meevasana,Pan,Chang,Inosov} as well as understanding their origin and implications\cite{Byczuk,QHWang,TZhou,Macridin,Markiewicz,Zhu,Manousakis,Alexandrov,Zemljic,Srivastava,FTan,Leigh}.
Along the nodal direction, angle-resolved photoemission (ARPES) measurements have revealed a high energy ``kink" feature near 400 meV and a high energy band (``waterfall" features) between -0.4 eV and $\sim$-1.0 eV\cite{Ronning,Graf,Xie,Valla,Meevasana,Pan,Chang,Inosov}. The origin of these high-energy features is under hot debate\cite{Ronning,Graf,Xie,Valla,Meevasana,Pan,Chang,Inosov,Byczuk,QHWang,TZhou,Macridin,Markiewicz,Zhu,Manousakis,Alexandrov,Zemljic,Srivastava,FTan,Leigh}
and it remains unclear whether they represent intrinsic band structure\cite{Inosov}. The clarification of these issues is important in establishing a basic theoretical framework to describe strongly correlated electron systems like cuprates, in probing electron dynamics by extracting electron self-energy, and in unraveling possible new physics.

\begin{figure}[tbp]
\begin{center}
\includegraphics[width=0.9\columnwidth,angle=0]{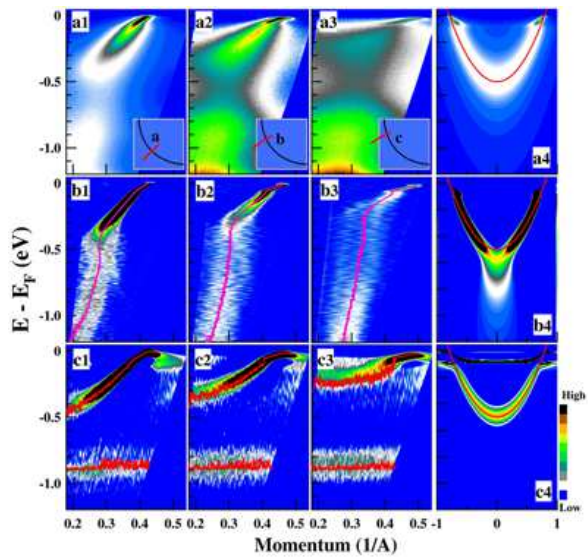}
\end{center}
\caption{Momentum dependent photoemission data(a1, a2 and a3) on optimally-doped Bi2212 measured at 17 K and their corresponding MDC second-derivative images (b1,b2 and b3) and  EDC second-derivative ones (c1, c2 and c3). The corresponding momentum cuts are indicated in insets for each image (a1-a3). The quantitatively derived MDC and EDC dispersions are overlaid in corresponding second-derivative images. (a4) shows a simulated single-particle spectral function for an electron-phonon coupling case. The bare band is a parabola (solid red line) and the phonon frequency cutoff is set at 70 meV. The corresponding MDC and EDC second-derivatives are shown in (b4) and (c4), respectively.
}
\end{figure}

\begin{figure*}[floatfix]
\begin{center}
\includegraphics[width=1.8\columnwidth,angle=0]{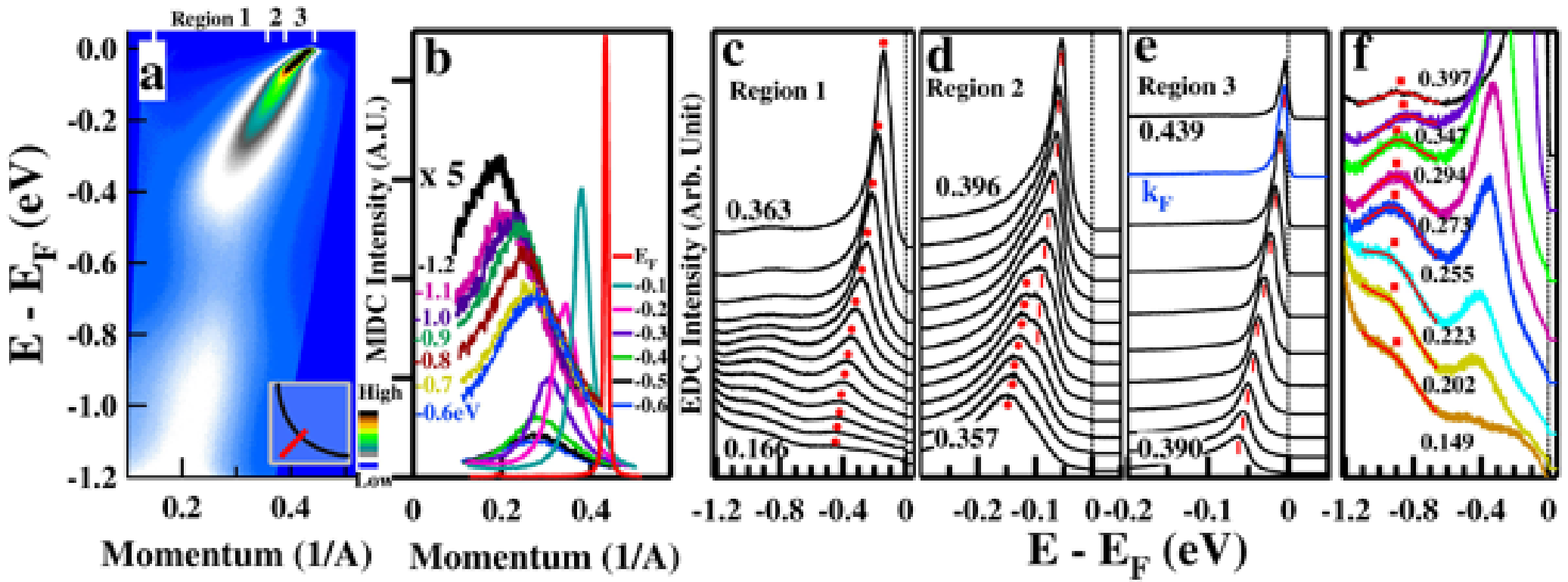}
\end{center}
\caption{(a). Photoemission data of Bi2212 measured along the $\Gamma$(0,0)-Y($\pi$,$\pi$) nodal direction at 17 K.  (b). Representative MDCs at various binding energies. For clarity, the MDCs between -0.6 and -1.2 eV are multiplied by 5. (c-e). Corresponding EDCs at different momentum regions as marked on the top of Fig. 2a. (f). Selected EDCs covering the entire region to highlight the broad peaks around -0.9 eV that are fitted by Lorentzians and marked by red solid squares.
}
\end{figure*}

In this paper, we examine the intrinsic band structure and the nature of the high energy band in Bi$_2$Sr$_2$CaCu$_2$O$_8$ (Bi2212) superconductor by taking advantage of the laser-based ARPES\cite{LiuIOP}. Our detailed momentum-dependent measurements clearly indicate that the MDC(momentum distribution curve)-derived high energy bands do not represent  LDA calculated ones and the high energy kink is unlikely due to electron coupling with high energy modes. The combined MDC and EDC (energy distribution curve) analyzes provide a complete band structure picture of Bi2212. The dichotomy between the MDC and EDC analysis raises a general question about its origin and which one represents the intrinsic band structure.  We find that the high energy MDC features are strongly affected by the existence of  high-intensity valence band and may not represent intrinsic band structure of Bi2212.

The ARPES measurements were carried out on our newly-developed vacuum ultraviolet (VUV) laser-based system\cite{LiuIOP}. The photon energy is 6.994 eV with a band width of 0.26 meV. The energy resolution of the electron energy analyzer (Scienta R4000) is set at 0.5 meV, giving rise to an overall energy resolution of 0.56 meV.  The momentum resolution is $\sim$0.004 $\AA$$^{-1}$. The photon flux is adjusted between 10$^{13}$ and 10$^{14}$ photons/second. The Bi2212 single crystals were cleaved {\it in situ} in vacuum with a base pressure better than 5$\times$10$^{-11}$ Torr.

Fig. 1(a1-a3) show momentum-dependent photoemission data of an optimally-doped Bi2212 (T$_c$=91K) measured at 17 K\cite{AntiFS}.  The two-dimensional images allow two different ways in extracting the corresponding band structure: the EDC and MDC analyzes. In Fig. 1 we extract the MDC band (Fig. 1(b1-b3)) and EDC band (Fig. 1(c1-c3)) by taking the second-derivative of the original photoemission data (Fig. 1(a1-a3)) with respect to momentum and energy, respectively. In this empirical way, the high-intensity contours on the second-derivative images can be related with possible band dispersions, as shown by their good agreement with the overlaid  MDC and EDC dispersions extracted quantitatively, as will be elaborated below.

\begin{figure}[b]
\begin{center}
\includegraphics[width=0.95\linewidth,angle=0]{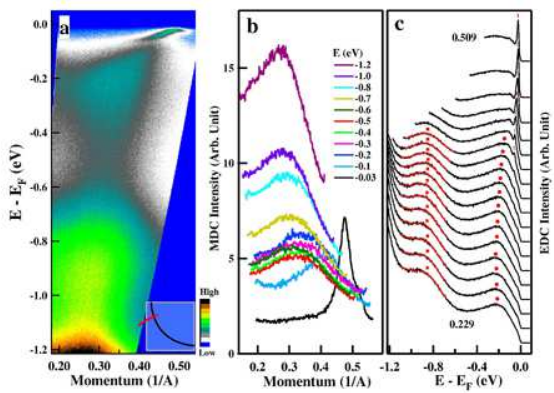}
\end{center}
\caption{(a). Photoemission image of Bi2212 measured at 17 K along the momentum cut {\it c} as indicated in the inset. (b). The corresponding MDCs at various binding energies. (c). The corresponding EDCs at various momenta. The peaks around -0.9 eV are fitted by Lorentzians and marked by red solid squares.
}
\end{figure}

As seen from Fig. 1, there is a dramatic difference between the MDC- and EDC-derived band structure.  The MDC-derived bands (Fig. 1(b1-b3)) give an overall dispersion with two obvious kink features, one at low energy near 50$\sim$70 meV that is well studied in the literature\cite{ThreeReviews} and will not be discussed in the present paper, the other is near 400 meV along the nodal direction (cut a, Fig. 1b1) which shifts to lower binding energy when the momentum cuts are moving to antinodal region: $\sim$310 meV for the cut {\it b} (Fig. 1b2) and $\sim$230 meV for the cut c (Fig. 1b3). No signature of -0.8 eV energy scale, as reported before\cite{Graf}, is observed. The strong momentum dependence of the high energy kink in Bi2212 is different from that in Bi2201 where it shows little change with momentum and can even be scaled\cite{Xie}.  On the other hand, the EDC-derived bands (Fig. 1(c1-c3)) are composed of two main branches. The first is the pronounced low energy band with a width of $\sim$0.5 eV for the nodal direction (Fig. 1c1); it gets narrower when the cuts are moving to antinodal region (Fig. 1c2 and 1c3).  The other branch is the high energy band near -0.9 eV that shows little momentum dependence. This band is not observed in previous measurements\cite{Graf,Valla,Meevasana,Inosov} but is similar to the -0.8 eV one recently identified in Bi2212 under special conditions\cite{Ding}. It is considered to be due to the interstitial oxygens in the Bi$_2$O$_2$ layers\cite{Ding}. We note that such a flat band can also be present in the t-J model calculations\cite{FTan} although it is not present in the LDA band calculations\cite{LDABand}.

Fig. 2 and Fig. 3 show individual MDCs and EDCs for the two typical cuts {\it a} and {\it c}, respectively.  The much improved resolution has resulted in sharper features that facilitate EDC analysis of band dispersion. Along the nodal direction (Fig. 2), one can see extremely sharp quasiparticle peaks near the Fermi momentum (region 3, Fig. 2e), coexistence of the peak and the emergent second peak (region 2, Fig. 2d) and dispersion of the second peak to high energy approaching $\sim$-0.5 eV (region 1, Fig. 2c). Such a detailed EDC evolution gives rise to the EDC dispersion shown in Fig. 4a from which one can also see low energy kink clearly near 70 meV. As seen from Fig. 2f, there are broad but well-defined peaks at high binding energy near -0.9 eV with little momentum dependence.  The EDCs (Fig. 3c) near the antinodal region (cut c) exhibit qualitatively similar behavior to the nodal cut and the EDC-derived dispersions are shown in Fig. 4c.

Fig. 4(a-c) presents a direct comparison between the MDC- and EDC-derived dispersions along the three different cuts. The corresponding LDA calculated bands are also included for comparison\cite{LDABand}.  It is clear that within the lower kink energy range (0$\sim$70 meV), the MDC- and EDC-derived dispersions show nearly perfect agreement.  Between the lower energy and high energy kinks, the MDC- and EDC-derived dispersions still show reasonably good agreement although the deviation gets larger with increasing binding energy. The dramatic discrepancy occurs above the high energy kink where the MDC- and EDC-derived bands show nearly no similarity.

It was proposed that the high energy MDC-derived dispersion might represent the recovery of the LDA-calculated band\cite{Ronning,Graf,Xie,Meevasana,Valla}. If this proves to be true, it would be significant as it provides a genuine bare band for extracting electron self-energy. As shown in Fig. 4a, if one focuses only on the nodal direction, it seems that the nodal MDC band is similar in band width to the LDA calculated one\cite{Markiewicz}. However, for the cuts (Fig. 4b-c) away from the nodal direction, the LDA calculated bands deviate significantly from the measured MDC dispersions.  In fact, there are disagreements even for the nodal direction\cite{Ronning,Graf,Xie,Meevasana}. Therefore, the momentum dependent measurements indicate that the consistency between the MDC-derived bands and LDA-calculated ones is not a general observation.

Another key issue is on the origin of the high energy kink near 200$\sim$400 meV in the MDC-derived dispersions. First, we note that such a high energy kink can not be generated from a strong electron coupling with low energy modes like phonons, even in polaronic regime\cite{Xie}. This has been experimentally demonstrated in a typical strong-coupling manganite system where no high energy kink is present\cite{MannellaNature,Meevasana}. The next question then comes  specifically to whether it can be attributed to the electron coupling with some high energy modes\cite{Valla}.  If one only focuses on a given momentum cut,  it is possible to make the simulated band match the measured MDC-derived one by tuning ``bare band", related bosonic spectral function, and other coupling parameters\cite{Valla}. However, the major problem with this scenario is that it is unable to provide a consistent description on the momentum dependence of the high energy MDC bands. First, it is hard to account for the evolution of the high energy kink from $\sim$400 meV (cut a) to $\sim$310 meV (cut b) and then to $\sim$230 meV (cut c) because this would ask for the existence of peak structures in the underlying bosonic spectral function at these specific energies.  It is more problematic that, in order to fit the MDC-derived high energy bands, one has to use the ``bare bands" which get steeper and wider for the cuts from nodal to antinodal regions, which is apparently not reasonable.   Therefore, the detailed momentum dependent measurements makes it clear that it is unlikely to attribute the high energy kink to the electron coupling with high energy modes.

As EDC and MDC analyzes are the two most popular ways in analyzing photoemission data, the dichotomy between the MDC- and EDC-derived bands from the same data raises critical questions about its origin and which one represents intrinsic band structure. These are general to the ARPES technique that has been a powerful tool in studying high-T$_c$ superconductors and other materials. In a typical Fermi liquid picture, the MDC- and EDC-derived dispersions are identical. In an electron-boson coupling system, except right near the kink region,  the lower and higher energy regions of the MDC- and EDC-derived dispersions are still consistent. These can be demonstrated from simulations and are experimentally proven even for strong electron-phonon coupling case\cite{MannellaNature}. The good agreement below the lower kink energy (E$_F$$\sim$ 70 meV) (Fig. 4a-c) is consistent with this general picture. The increasing deviation above the lower kink energy (70$\sim$500 meV) suggests some new factors coming into play, with possibilities like strong correlation, incoherence or momentum dependent electron self-energy.
The dramatic disparity between the MDC and EDC analyzes on the high energy dispersions (-0.5$\sim$-1.2 eV) points to its unusual nature.

\begin{figure}[tbp]
\begin{center}
\includegraphics[width=1.00\columnwidth,angle=0]{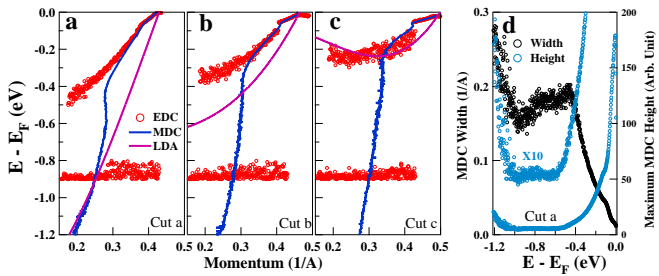}
\end{center}
\caption{(a-c).Comparison between the MDC-derived (blue lines)  and  EDC-derived dispersions (red circles) for the three momentum cuts. The corresponding LDA calculated bands\cite{LDABand} are also plotted (purple lines). (d). The fitted MDC width and height for the nodal cut {\it a}. For clarity of the high energy part, the MDC height data is also multiplied by 10.
}
\end{figure}

Generally, peaks in EDCs are robust signatures in representing intrinsic band structure, but the MDC analysis, in spite of its advantages over EDC analysis in some respects, should be used with caution. This is because the MDC analysis is applicable under some assumptions and it may be complicated by spurious artifacts, particularly near the bottom or top of a band, near localized strong intensity patch, for multiple bands or in the gapped region\cite{NormanMDC}.  As an illustration, Fig. 1a4 shows a simulated image of an electron-phonon coupling case with a parabolic bare band. While the EDC analysis gives a right band structure at high energy(Fig. 1b4), the MDC analysis deviates near the band bottom and even produces a nearly vertical dispersion at high energy below the band bottom that is obviously an artifact.  In Bi2212, we notice that the EDCs (Figs. 2f and 3c) at high binding energy are sitting on a rising background, leading to corresponding MDC intensity increase with increasing binding energy(Figs. 2b and 3b). This background is most likely from the tail of the adjacent strong valence band that shoots up between -1.0 and -1.2 eV (Fig. 2f and 3c).  In particular, the MDCs at -1.2 eV (Fig. 2b and 3b), which are already inside the strong valence band, still exhibit well-defined peaks. The resultant MDC width and MDC peak height (Fig. 4d) between -1.2 eV and -1.0 eV  sense the existence of the strong valence band but the corresponding MDC dispersion (Fig. 4a) does not exhibit an obvious change near -1.0 eV.  These observations suggest that the high energy MDC feature between -0.2 and -1.0 eV may be closely connected to the adjacent valence band near -1.0$\sim$-1.2 eV.  Because it is common to have a strong intensity patch of valence band at high binding energy beyond  -1.0 eV\cite{Ronning,Graf,Meevasana,Inosov}, it is possible to induce tailing effect at low binding energy that mimics an MDC dispersion.  Such an effect must play an important role although it needs to be further explored whether it can fully account for the high energy MDC dispersion.

In summary, from our super-high resolution and momentum dependent measurements we have gained new insights on the intrinsic band structure and nature of the MDC high energy feature in Bi2212. Combining MDC and EDC analyzes is important in arriving at a complete band structure and further theoretical work needs to be done to understand their dichotomy. The high energy MDC dispersion may be strongly affected by the existence of a strong valence band below -1.0 eV and may not represent intrinsic band structure.

We acknowledge helpful discussions with Q. H. Wang and J. R. Shi. This work is supported by the NSFC, the MOST of China (973 project No: 2006CB601002, 2006CB921302), and CAS (Projects ITSNEM and 100-Talent).  The work at BNL is supported by the DOE under contract No. DE-AC02-98CH10886.

$^{*}$Corresponding author (XJZhou@aphy.iphy.ac.cn)



\begin{thebibliography}{99}
\bibitem{AntiFS} At the photon energy of 6.994 eV, only the antibonding Fermi surface sheet is observed.
\end{thebibliography}
\end{document}